\documentclass[a4paper,12pt]{article}
\pdfoutput=1
\usepackage{a4wide}
\usepackage{amsfonts}
\usepackage{mathrsfs}
\usepackage{graphicx}
\usepackage{amssymb}
\usepackage{color}
\usepackage{amsmath}
\usepackage{epstopdf}
\begin{document}


%
%

\title{Understanding strongly coupling magnetism from holographic duality}

\author{Rong-Gen Cai\footnote{E-mail: cairg@itp.ac.cn} and Run-Qiu Yang\\
 Institute of Theoretical Physics, Chinese Academy of Sciences, 
\\Beijing 100190, China.}



\date{}
\maketitle

\begin{abstract}
The unusual magnetic materials are significant in both  science and technology. However, because of the strongly correlated effects, it is difficult to understand their novel properties from theoretical aspects. Holographic duality offers a new approach to understanding such systems from gravity side. This paper will give a brief review of our recent works on the applications of holographic duality in understanding unusual magnetic materials. Some quantitative compare between holographic results and experimental data will be shown and some predictions from holographic duality models will be discussed.
\end{abstract}




\section{Introduction}	
 The AdS/CFT correspondence or gauge/gravity duality, one of important progresses in quantum gravity,  has  offered a new viewpoint to understanding the connection between quantum gravity and conformal field theory~\cite{Maldacena:1997re,Gubser:1998bc,Witten:1998qj,Witten:1998zw}.  This correspondence relations a weakly coupling gravity in $(d+1)$-dimensional anti-de Sitter (AdS) space-time to a strongly coupling conformal field theory (CFT) in $d$-dimensional spacetime. As the correspondence maps a strongly coupling many-body problem into a weakly coupling gravity system in one more dimension, it opens a new approach to deal with the strongly correlated electron systems which are difficult to handle in traditional condensed matter theories. Over the past years, this correspondence has been applied into an abundant of strongly correlated condensed systems such as  superconductor/superfuild~\cite{Hartnoll:2008vx,Hartnoll:2008kx,Hartnoll:2009sz,Herzog:2009xv,McGreevy:2009xe,Horowitz:2010gk,Cai:2015cya}, fermi/non-fermi liquids~\cite{Lee:2008xf,Liu:2009dm,Cubrovic:2009ye}, charge density wave and metal/insulator phase transition~\cite{Aperis:2010cd,Donos:2013gda,Ling:2014saa} and some systems far from thermal equilibrium~\cite{Murata:2010dx,Bhaseen:2012gg,Adams:2012pj,Garcia-Garcia:2013rha,Chesler:2013lia}.  In these works, the attentions are focused on the electric transport properties in strongly correlated materials such as conductivity. Though there are a few of works involving magnetism in  holographic superconductor models, magnetism does not pay the central role. However, some important strongly correlated phenomena  including high temperature superconductors and heavy fermion metals are controlled by magnetism of the material. Because of the strong correlation between electrons in the materials, there is still lack of a well theoretical description.  The gauge/gravity duality provides  an approach and perspective to understand these challenging problems.

As the spontaneous U(1) symmetry breaking, which  plays am important role in building holographic superconductor models, the time reversal symmetry breaking is the characteristic of magnetic ordered states. An  important property in magnetic materials is the response to external magnetic field. In  phenomenology, a holographic model which can describe the phenomena involving spontaneous magnetic ordered state should give time reversal symmetry breaking spontaneously and characteristic response to external magnetic field. In this paper, we will introduce  our recent works about using holographic duality to understand strong coupling phenomena controlled by magnetism. We will first give  a holographic model to realize the ferromagnetic phase transition and discuss its important properties. Then we will show some singificant results in understanding some novel magnetic materials based on this model, including the coexistence and competition between ferromagnetism and superconductivity, colossal magnetic resistance and antiferromagnetic quantum phase transition induced by an external magnetic field.

\section{Holographic ferromagnetic model}
The model we are considering is Einstein-Maxwell theory in asymptotic AdS space-time. The first holographic ferromagnetic model was proposed in Ref.~\cite{Cai:2014oca}. However, this mode exists external degree of freedoms which leads ghost or causality violation. Here we will introduce a modified model proposed in Ref.~\cite{Cai:2015bsa}. The action in the bulk  is,
\begin{eqnarray}\label{action1}
S&=&\frac{1}{2\kappa^2}\int d^4x\sqrt{-g}(L_{1}+\lambda^2 L_{2}),\nonumber\\
L_{1}&=&\mathcal{R}+\frac{6}{L^2}-F^{\mu\nu}F_{\mu\nu},\\
L_{2}&=&-\frac{1}{12}(d M)^2-\frac{m^2}{4}M_{\mu\nu}M^{\mu\nu}-M^{\mu\nu}F_{\mu\nu}-\frac{J}{8}V(M).\nonumber
\end{eqnarray}
where $L$ is the AdS radius and $2\kappa^2=16\pi G$ is associated with the gravitational constant in the bulk. We will set them both to be unity in the following. $g$ is the determinant of the bulk metric $g_{\mu\nu}$. $A_{\mu}$ is U(1) gauge potential and $F_{\mu\nu}=(dA)_{\mu\nu}$.  $d M$ is the exterior differential of 2-form field $M_{\mu\nu}$. $m^2, \lambda$ and $J$ are three real model parameters with $J<0$ and $m^2>0$ in order the magnetization to happen spontaneously. $V(M_{\mu\nu})$ is a nonlinear potential of the 2-form field to describe the self-interaction of the polarization tensor. It should be expanded as the even power of $M_{\mu\nu}$. The choice of nonlinear potential is not unique. Following Ref.~\cite{Cai:2015bsa}, we take the following form,
\begin{eqnarray}\label{expV}
V(M_{\mu\nu})&=&(^{*}M_{\mu\nu}M^{\mu\nu})^2=[^{*}(M\wedge M)]^2.
\end{eqnarray}
Here $^{*}$ is the Hodge-star operator.

In the simple case, the dual boundary is isotropy and homogenous, so all the fields in the bulk are only  functions of radius $r$. Specially, we will use following ansatz for metric and matter field,
\begin{eqnarray}\label{5}
ds^2&=&-r^2 f(r) e^{a(r)}dt^2+\frac{dr^2}{r^2 f(r)}+r^2(dx^2+dy^2),
\end{eqnarray}
\begin{eqnarray}\label{6}
M_{\mu\nu}=-p(r)dt\wedge dr+\rho(r)dx\wedge dy,~~A_{\mu}&=&\phi(r)dt+Bx dy,
\end{eqnarray}
with some real functions $f(r)$, $a(r)$, $\phi(r)$, $p(r)$,and $\rho(r)$, which are determined by equations of motions through the variation of action~\eqref{action1}.

The bulk field $B$ is a constant magnetic field, which can be regarded as external magnetic field in the dual boundary theory. We will denote the position of the horizon as $r_{h}$ and the conformal boundary will be at $r\rightarrow\infty$. In order to study a dual theory with finite chemical potential or charge density accompanied by a U(1) symmetry, we need turn on $A_{t}=\phi(r)$ in the bulk. The magnetic moment density in the dual boundary is expressed as,
\begin{equation}\label{defN1}
N=-\frac{\lambda^2}{2}\int_{r_h}^\infty\frac{\rho e^{a/2}}{r^2}dr.
\end{equation}
It has been shown that\cite{Cai:2015jta} $\rho\rightarrow-\rho$ as $B\rightarrow-B$ under the time reversal transformation, which leads to the property of magnetic moment defined in Eq.~\eqref{defN1} under the time reversal transformation such as $N\rightarrow-N$. This is agreement with the fact that magnetic field is pseudo-vector in 3+1 boundary dimensions or pseudo-scalar in 2+1 boundary dimensions. In the case of $B=0$, the appearance of nonzero $\rho$ will lead to spontaneous magnetization and time reversal symmetry breaking.

\begin{figure}
\centering
\includegraphics[width=0.4\textwidth]{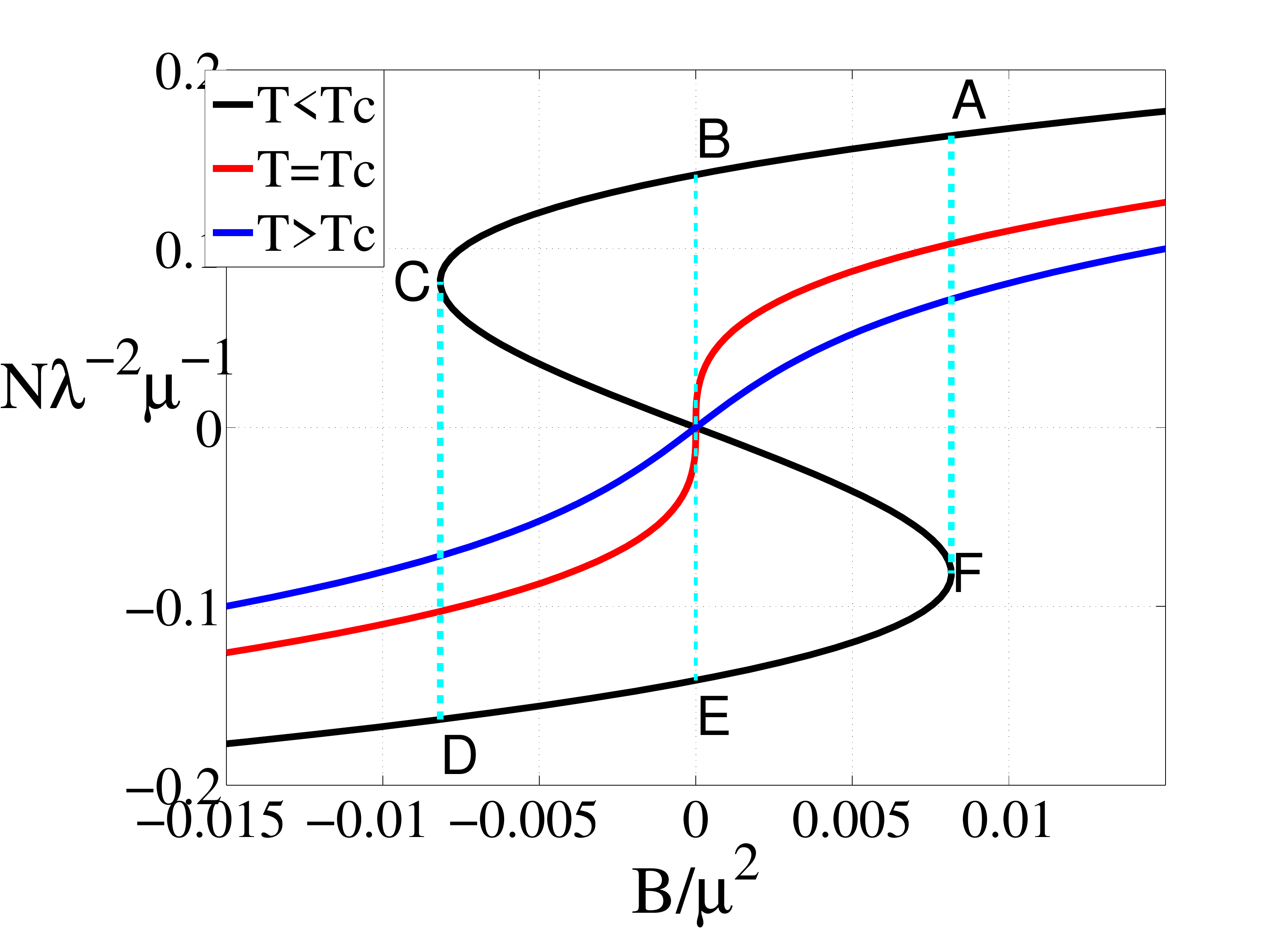}
\includegraphics[width=0.4\textwidth]{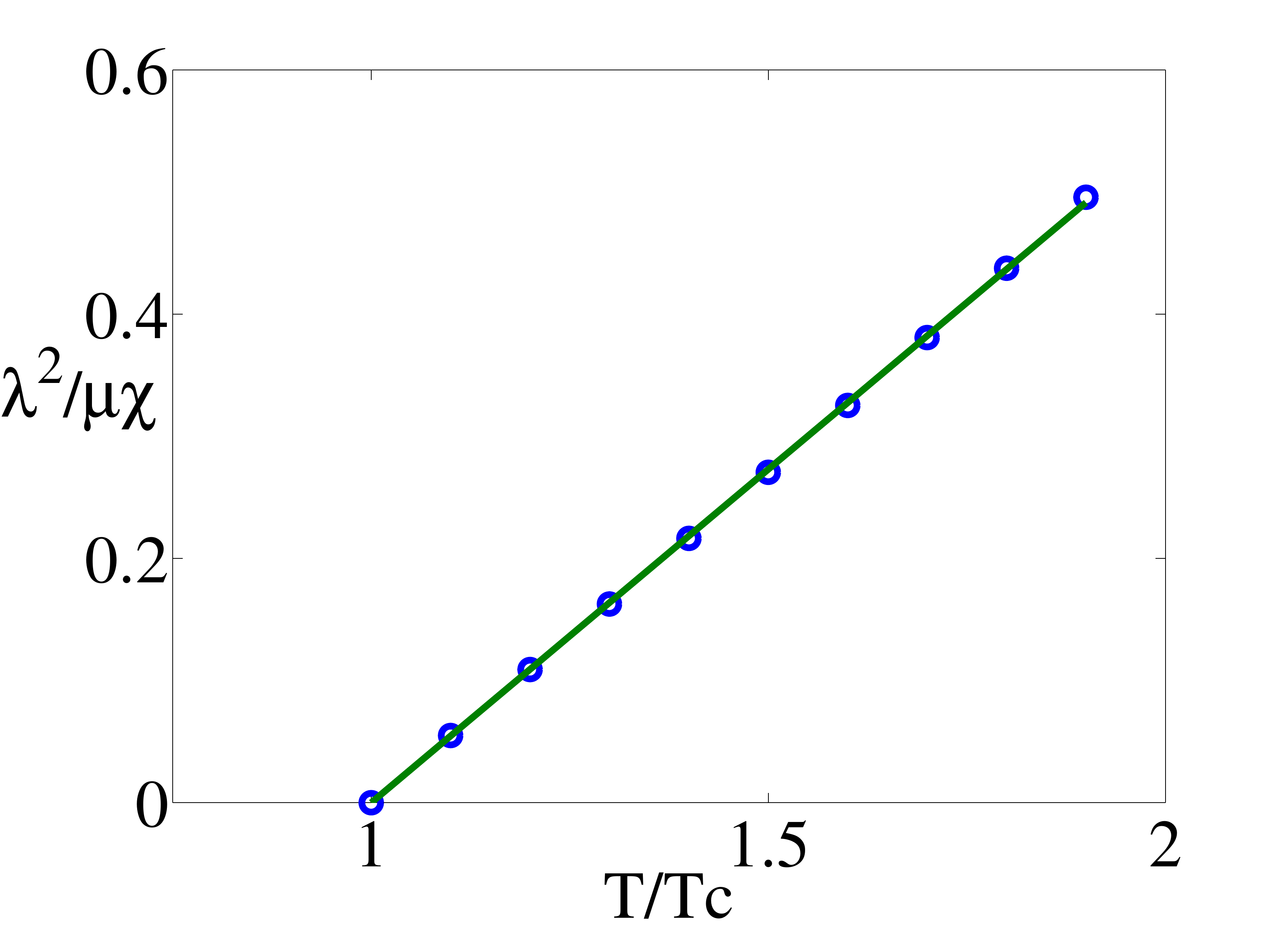}
\vspace*{8pt}
\caption{The magnetic moment $N$ (left panel) as a function of temperature in different magnetic field $B$ and the magnetic susceptibility $\chi$ (right panel) as a function of temperature. Here $\mu$ is the chemical potential of dual boundary.} \label{F1}
\end{figure}

The key physics of this model can be understood by the probe limit when $\lambda\rightarrow0$. In this limit, we  neglect the back reaction of polarization field to the gauge field and background geometry. In such a case, the background is just the dyonic Reissner-Nordstrom AdS black hole. In the vicinity of critical temperature and weak external magnetic field, the magnetic part of the grand thermodynamic potential density $\Omega$ in grand canonical ensemble is\cite{Cai:2015jta},
\begin{equation}\label{expand4}
\Omega\simeq -BN+\frac{2a_0}{\lambda^2N_1^2}(T/T_c-1)N^2+\frac{-16\widetilde{J}_fa_1}{\lambda^6N_1^4}N^4+\mathcal{O}(N^6).
\end{equation}
Here $a_0, a_1>0, N_1$ and  $\tilde{J}_f<0$ are all constants, the details of these constants can be found in Ref.~\cite{Cai:2015jta}. We can see that this is just the Ginzburg-Landau (GL) theory of ferromagnetic model. From it, we can obtain the expression of magnetic moment in the ferromagnetic phase when $B=0$,
\begin{equation}\label{onshellN}
N/\lambda^2=\sqrt{\frac{N_1^2a_0}{-16\widetilde{J}_fa_1}}(1-T/T_c)^{1/2}, ~~\text{when}~T\rightarrow T_c^-.
\end{equation}
This tells us that the critical exponent $1/2$ is an exact result.  In the nonzero magnetic field case, the expression \eqref{expand4} gives out correct magnetic susceptibility behavior and hysteresis loop. When the temperature is far way from its critical value, then the expression \eqref{expand4} is not appliable. The spontaneous magnetic moment density and magnetic susceptibility can be determined numerically from the equations of motion. One typical result is shown in  Fig.~\ref{F1}. One can find that the model realizes the hysteresis loop of single magnetic domain and that the magnetic susceptibility satisfies the Curie-Weiss law.

The case with full back reaction has also be studied in Ref.~\cite{Cai:2015jta}, which shows that ferromagnetic phase transition will also happen at low temperature and the phase transition is always a second order one when we increase the strength of the back reaction.

\begin{figure}
\centering
\includegraphics[width=0.38\textwidth]{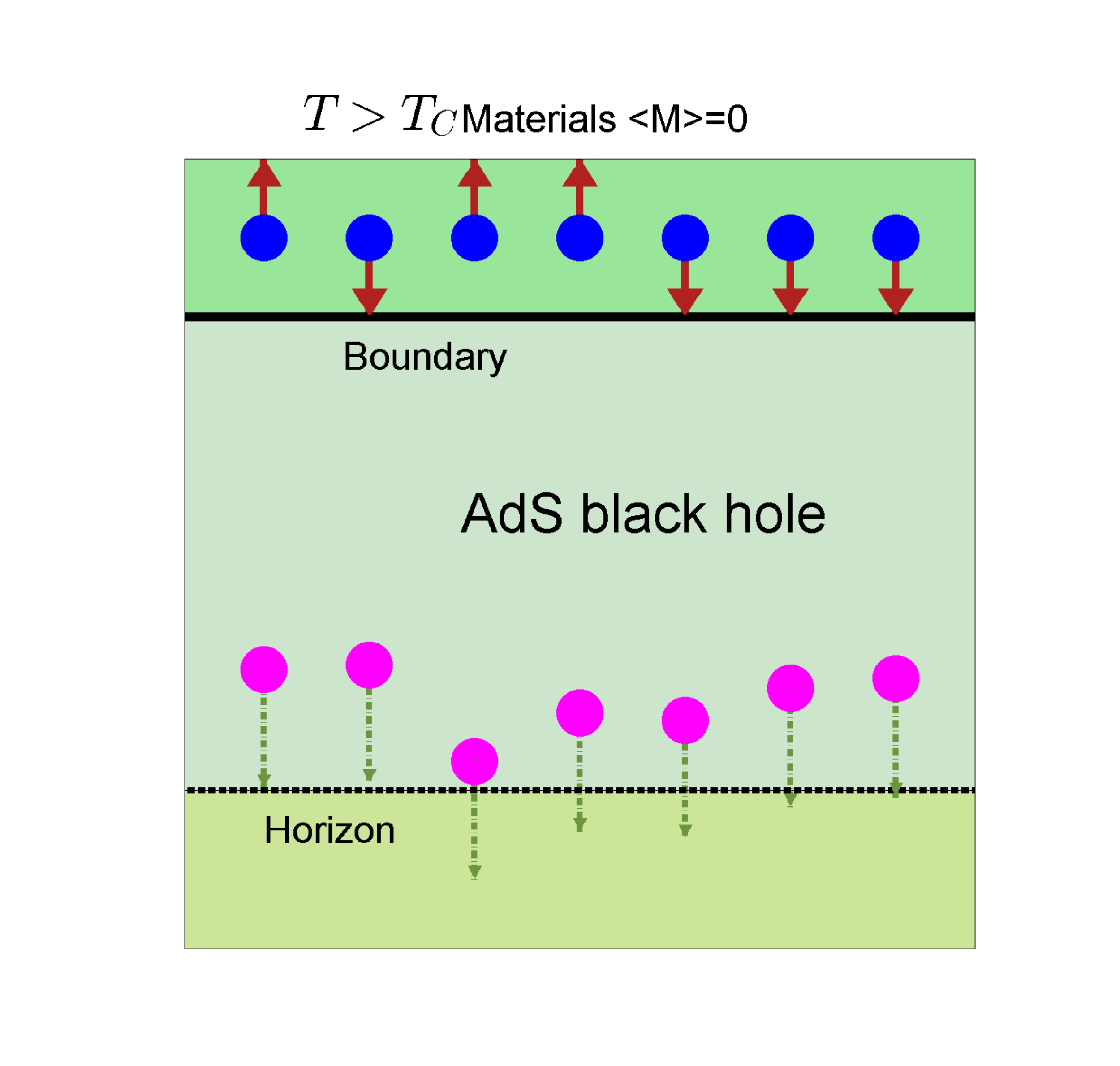}
\includegraphics[width=0.4\textwidth]{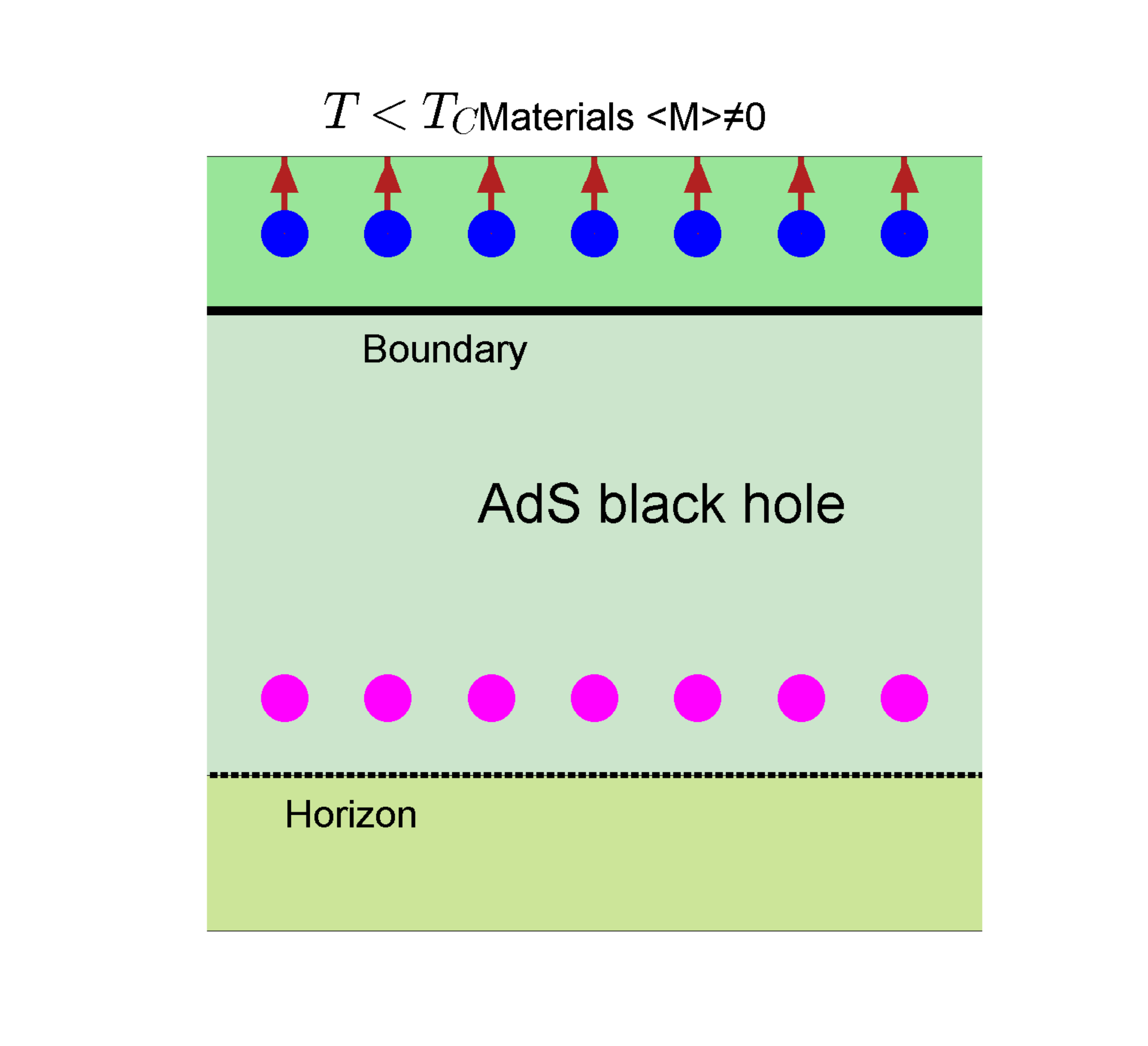}
\vspace*{8pt}
\caption{The schematic diagram for the condensation of the 2-form field near the black hole horizon. The pink circles stand for the  condensation of the spatial part of 2-form field caused by vacuum fluctuation. The blue circles stand for  microcosmic magnetic moments in materials and the arrows stand for their directions. $T$ is  temperature and $T_C$ is the critical temperature. } \label{F2}
\end{figure}

 The holographic ferromagnetic phase transition is a consequence of the repulsion of the 2-form field and the attraction from gravity. This can be understood from Fig.~\ref{F2}  schematically. As the high temperature corresponds to strong surface gravity and the gravity dominates the system, the condensation of the  2-form field caused by vacuum fluctuation will be attracted into the black hole and disappear. This implies that the thermal fluctuation
dominates the materials in high temperature and the ordered magnetic moment will become disordered quickly because of thermal fluctuation. In low temperature, the surface gravity becomes weak and the repulsion dominate the system.  The condensation of the 2-form field caused by vacuum fluctuation can be held near the horizon. This implies that the thermal fluctuation is weak and cannot destroy the the ordered magnetic moment so  that the macroscopical magnetic moment appears.

\section{Application in strongly correlated materials}
Though ferromagnetic phase transition itself has been understood well in condensed matter theory, the holographic model offers a framework to investigate the strongly correlated phenomena controlled by magnetism. As this framework is independent of the model such as holographic superconductor or fermi/non-fermi liquid, we can combine it with them and study spontaneous magnetic order in unconventional superconductors, strange metals or other novel materials.

\subsection{Competition and coexistence of  ferromagnetism and superconductivity}
One of interesting problems is the coexistence of  ferromagnetism and superconductivity in heavy fermi materials such as UGe$_2$\cite{Lonzarich},
URhGe\cite{Aoki} and UCoGe\cite{Huy}. In a very long time, it was thought these two phenomenons are incompatible with each other. In the viewpoint of BCS theory, any magnetic impurity will suppresses the singlet Cooper pair formation, which causes a rapid depression of the superconducting transition temperature. Likewise, the superconductivity can screen off the magnetic field and leads to  that the long-range magnetic order is accompanied by the expulsion of superconductivity. However, the new novel materials challenge this belief. In some  heavy fermi materials, the ferromagnetism and superconductivity are found to coexist with each other. The nature of this superconducting state in ferromagnetic materials is currently under debate. The heavy effective mass of  valence electrons shows the strong correlation between them, which implies that approximations and conception of free field lose their validness.

In Ref. \cite{Cai:2014dza}, the authors first study the competition and coexistence of  ferromagnetism and superconductivity from holographic duality. By  combining  a holographic $p$-wave superconductor model\cite{Cai:2013aca} and a holographic  ferromagnetism model, Ref. \cite{Cai:2014dza} studied the coexistence and competition of ferromagnetism and $p$-wave superconductivity. It is found that the results depend on the self-interaction of magnetic moment of the complex vector field and which phase appears first. In the case that the ferromagnetic phase appears first, if the interaction is attractive, the system  shows the ferromagnetism and superconductivity can coexist in low temperatures.  If the interaction is repulsive,  the system will only be in a pure ferromagnetic state.  In the case that the superconducting phase appears first, the attractive interaction will lead to a magnetic $p$-wave superconducting phase in low temperatures. If the interaction is repulsive, the system will transform into a pure ferromagnetic phase when the temperature is lowered.

\subsection{Colossal magnetic resistance in holographic duality}
Another interesting application is about the phenomena called ``colossal magnetoresistance" (CMR) effect, which appaers in the manganites such as A$_{1-x}$B$_x$MnO$_3$ (A= La, Pr, Sm, etc. and B = Ca, Sr, Ba, Pb). Such phenomena is among the main areas of research in strongly correlated electron systems~\cite{Urushibara,Uehara,MBS,Dagottoa,Nagaev,Cengiz,Mukherjee}. In these materials, there is remarkable magnetoresistivity near the Curie temperature $T_C$ and an insulator (or semiconductor)/metal phase transition associated with a paramagnetic/ferromagnetic phase transition happens. The CMR effect has a completely different physical origin from the ``giant" magnetoresistance observed in layered and clustered compounds. Because of the progress in computation and mean field studies for realistic models, many important results have been achieved. However, a completely understanding of the CMR effect is still a challenge.

\begin{figure}
\centering
\includegraphics[width=0.37\textwidth]{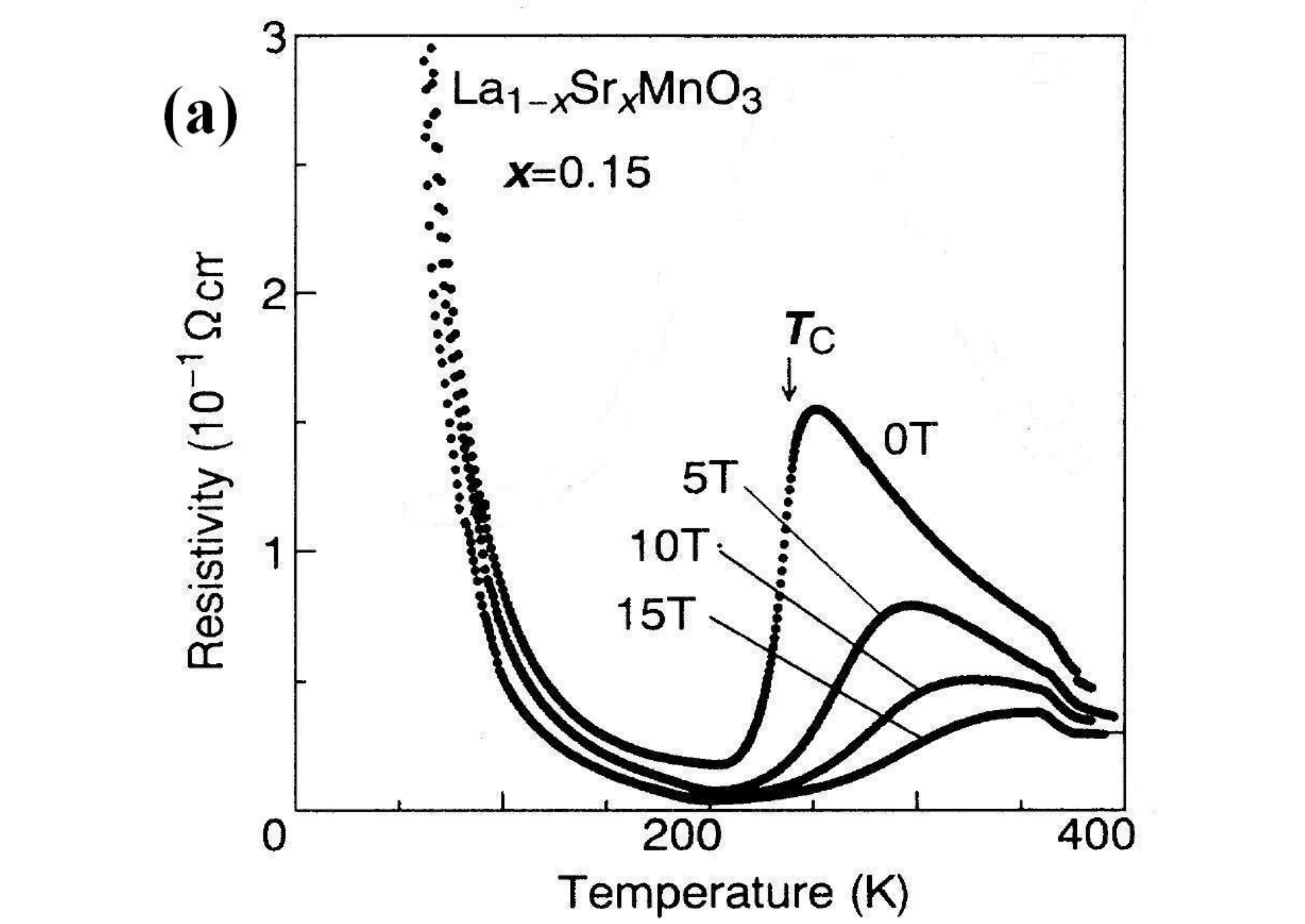}
\includegraphics[width=0.38\textwidth]{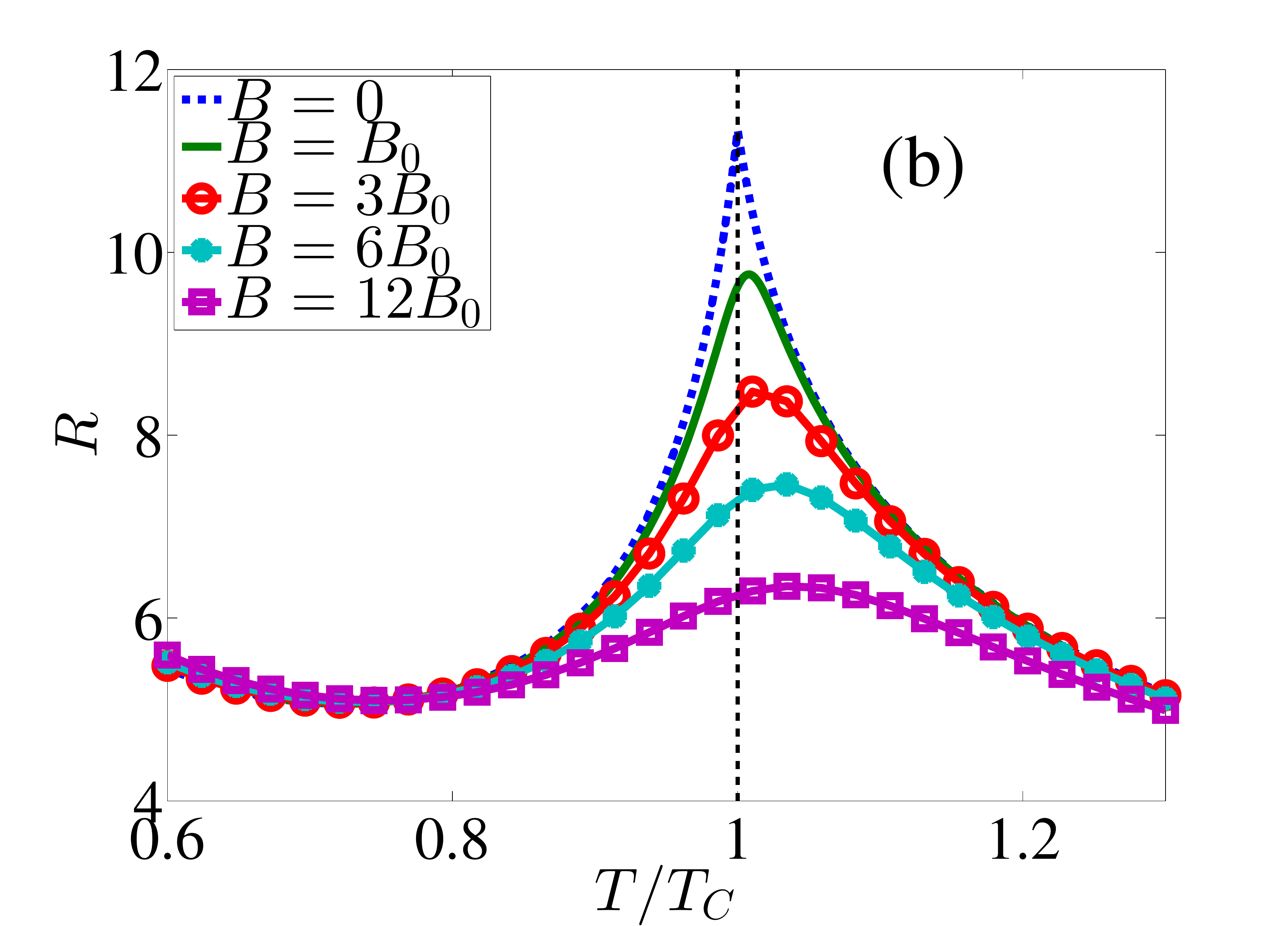}
\vspace*{8pt}
\caption{Left panel: the DC resistivity of La$_{1-x}$Sr$_x$MnO$_3$ as a function of temperature for different magnetic field~\cite{Urushibara}.   Right panel: the DC resistivity vs temperature in different external magnetic field.} \label{F3}
\end{figure}

Within massive gravity, Ref. \cite{Cai:2015wfa} constructs a gravity dual for insulator/metal phase transition and CMR effect. The main results in Ref. \cite{Cai:2015wfa} can be shown in  Fig.~\ref{F3}. From the right panel of Fig.~\ref{F3}, we see that the DC resistance shows an insulator behavior when the temperature is higher than $T_C$, while the temperature is less than the Curie temperature $T_C$, the resistivity decreases, which shows a metal's behavior. There is a distinct peak at the temperature where spontaneous magnetization begins to appear and an insulator/metal phase transition happens there.  This is just one of characteristic properties of CMR materials in manganese oxides.  What's more, when a small magnetic field $B$ is turned on, we find that the resistivity is very sensitive to the external magnetic field.
In addition, holographic model shows that magnetoresistance (MR) in the case of weak magnetic field and $T\rightarrow T_C^+$ obeys following relation,
\begin{equation}\label{scal1}
\text{MR}=1-R(B)/R(B=0)\propto N^2.
\end{equation}
It is in agreement with the experimental and theoretical analysis of CMR materials~\cite{Urushibara}.

\subsection{Antiferromagnetism and quantum phase transition induced by external magnetic field}
The holographic ferromagnetic model has been generalized so that it can describe antiferromagnetism phase transition in Ref.~\cite{Cai:2014jta}. In  the antiferromagnetic model, in order to describe the magnetic structures of antiferromagnetism, Ref. \cite{Cai:2014jta} introduces two different antisymmetry polarization fields in the bulk. These polarizations couple to U(1) gauge field strength and each other. The system shows a critical temperature $T_N$ if the values of parameters are in some suitable region.  In the case of absence of external magnetic field and low temperature, the magnetic moments of these tensor condense in antiparallel manner with the same magnitude, which leads to an antiferromagnetic state. In the weak external magnetic field, the magnetic susceptibility density has a peak at critical temperature and satisfies the Curie-Weiss law in the paramagnetic phase of antiferromagnetic materials.

One unexpected discovery is that the model in Ref.\cite{Cai:2014jta} shows critical temperature $T_N$ will be suppressed by external magnetic field. There is a critical magnetic field $B_c$ where the critical temperature will be suppressed into zero. This shows that this holographic antiferromagnetic model can describe a quantum phase transition (QPT) induced by magnetic field. Inspired by this result, Refs. \cite{Cai:2015mja,Cai:2015xpa} investigate a holographic model for antiferromagnetic quantum phase transition induced by magnetic field.
\begin{figure}
\centering
\includegraphics[width=0.35\textwidth]{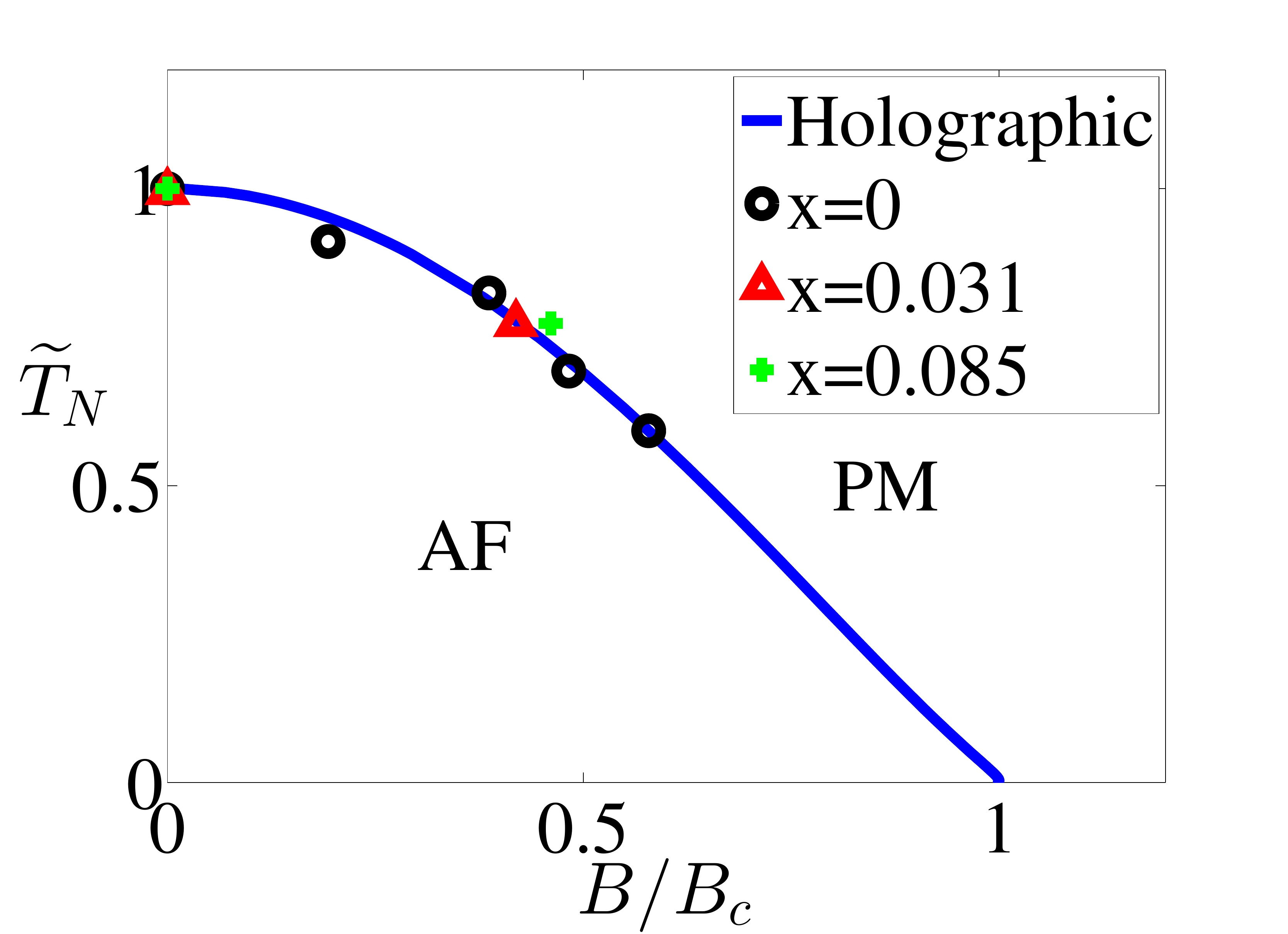}
\includegraphics[width=0.34\textwidth]{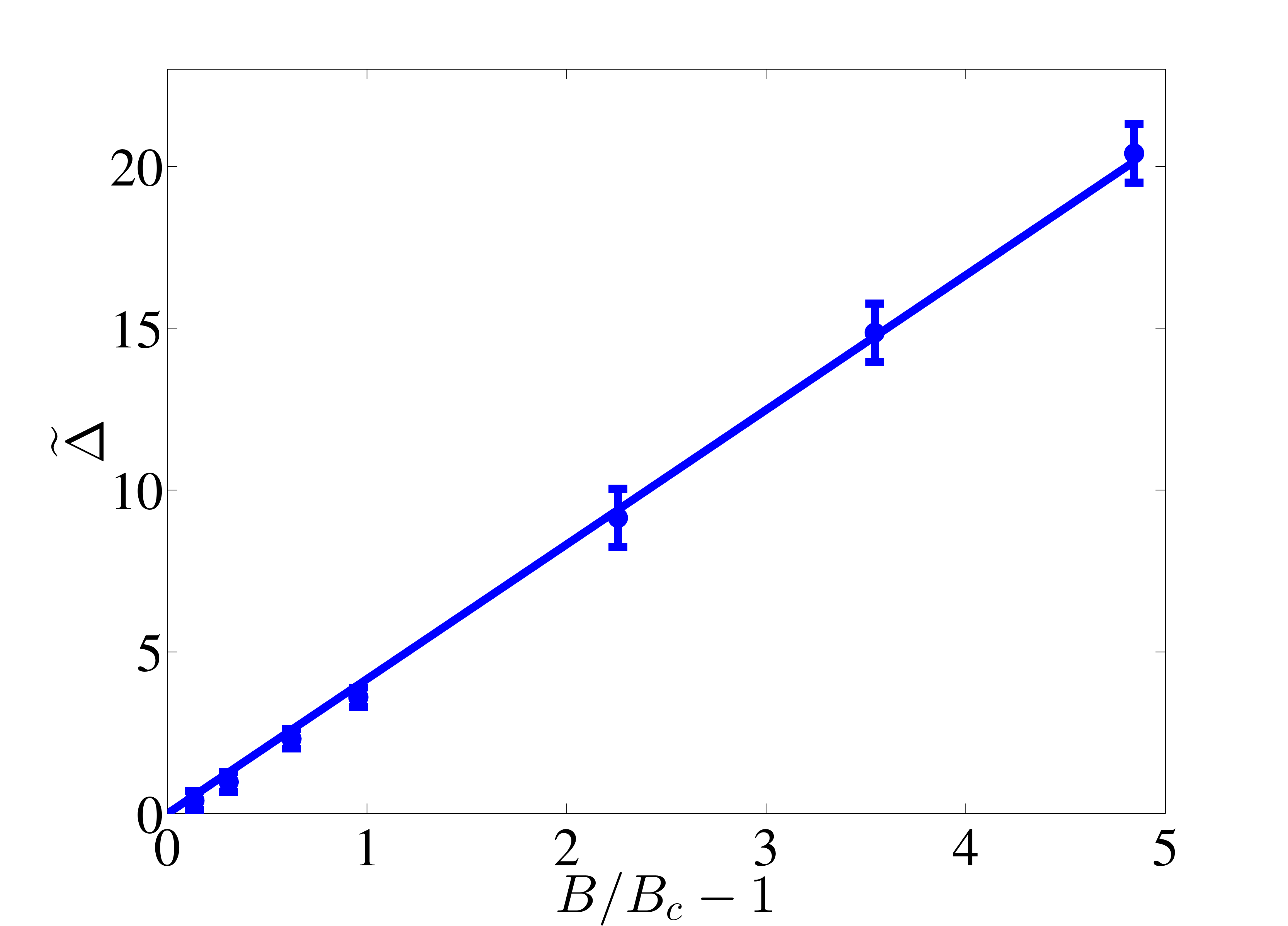}
\vspace*{8pt}
\caption{Left panel: the antiferromagnetic critical temperature $T_N$ versus the external magnetic field $B$. The solid curve is holographic results and the different dots are experimental data\cite{Cai:2015mja,Niven} for pyrochlore compounds,  Er$_{2-2x}$Y$_{2x}$Ti$_2$O$_7$ with three different doping $x$.   $T_N$ is the antiferromagnetic critical temperature and $T_{N0}$ is the value of $T_N$ at $B=0$. Right panel: the energy gap of the high-field ferrimagnetic state, $\Delta$, of Er$_{2-2x}$Y$_{2x}$Ti$_2$O$_7$ as a function of magnetic dilution ($x$) and magnetic field $B$ measured in the units of the critical field~\cite{Niven}. } \label{F4}
\end{figure}
 Fig.~\ref{F4} shows that results from holographic model and experimental results of QPT from  pyrochlores (Er$_{1-x}$Y$_x$)$_2$Ti$_2$O$_7$ ~\cite{Niven}. We can see that, after rescaling the  temperature and magnetic field, the antiferromagnetic transition boundary of (Er$_{1-x}$Y$_x$)$_2$Ti$_2$O$_7$ is  insensitive to the doping, which shows a universal behavior. The holographic model can cover well the experimental date. Besides, the holographic model forecasts two scaling relations,
\begin{eqnarray}
&\widetilde{\Delta}=C(B/B_c-1), \text{with}~\widetilde{\Delta}=\Delta/k_BT_{N0},\label{TB1a}\\
&\widetilde{T}_N/\ln \widetilde{T}_N\propto(1-B/B_c),\label{TB1b}
\end{eqnarray}
where $\widetilde{T}_N=T_N/T_{N0}$, $k_B$ is Boltzmann constant and $C$ is a constant. These scaling relations can be analytically presented by considering the emergent geometry AdS$_2$  in the  IR limit~\cite{Cai:2015xpa}. The first one has been measured in (Er$_{1-x}$Y$_x$)$_2$Ti$_2$O$_7$ (see the right panel of Fig.~\ref{F4}) and experimental data give constant $C\approx5.0$, which is very closed to the holographic result $C\approx4.2$. Because of the difficulty in experiments, it is still an open question for experimenter to check the other scaling relationship in Eq.\eqref{TB1b}.

\section{Summary}
Understanding magnetic properties of  materials has a very long history. However, it is still mysterious for lots of materials in extreme pressure and low temperature. Magnetism plays a crucial role in Mott transition, heavy fermion, and quantum phase transition. In fact, quantum magnetism is one of
the manifestations of strong electron-electron interaction, so its treatment should be part of general strongly correlated electron effects. Holographic duality is a very suitable tool for such phenomena. This paper gave a brief review about the applications of holographic duality in understanding unusual magnetic materials. Some quantitative comparison between holographic results and experimental data was made and some predictions from holographic duality models were discussed. Though some works have been done to understand the phenomena in strongly correlated electron systems controlled by magnetism, it is still far from a complete understanding. The most works at present stage are to  recover some properties which have been discovered in experiments. Of course, this is important and provides a starting point to build a holographic description for real materials.


\section*{Acknowledgements}
This work was supported in part by the National Natural Science Foundation of China  with grants No.11375247 and No.11435006.  We thank F. Kusmartsev, Y. B. Wu and
C.Y. Zhang for collaborations in some relevant studies.



\begin{thebibliography}{0}    

\bibitem{Maldacena:1997re}
  J.~M.~Maldacena,
  ``The Large N limit of superconformal field theories and supergravity,''
  Int.\ J.\ Theor.\ Phys.\  {\bf 38}, 1113 (1999)
  [Adv.\ Theor.\ Math.\ Phys.\  {\bf 2}, 231 (1998)]
  [hep-th/9711200].

\bibitem{Gubser:1998bc}
  S.~S.~Gubser, I.~R.~Klebanov and A.~M.~Polyakov,
  ``Gauge theory correlators from noncritical string theory,''
  Phys.\ Lett.\ B {\bf 428}, 105 (1998)
  [hep-th/9802109].

\bibitem{Witten:1998qj}
  E.~Witten,
  ``Anti-de Sitter space and holography,''
  Adv.\ Theor.\ Math.\ Phys.\  {\bf 2}, 253 (1998)
  [hep-th/9802150].

\bibitem{Witten:1998zw}
  E.~Witten,
  ``Anti-de Sitter space, thermal phase transition, and confinement in gauge theories,''
  Adv.\ Theor.\ Math.\ Phys.\  {\bf 2}, 505 (1998)
  [hep-th/9803131].


\bibitem{Hartnoll:2008vx}
  S.~A.~Hartnoll, C.~P.~Herzog and G.~T.~Horowitz,
  ``Building a Holographic Superconductor,''
  Phys.\ Rev.\ Lett.\  {\bf 101}, 031601 (2008)
  [arXiv:0803.3295 [hep-th]].

\bibitem{Hartnoll:2008kx}
  S.~A.~Hartnoll, C.~P.~Herzog and G.~T.~Horowitz,
  ``Holographic Superconductors,''
  JHEP {\bf 0812}, 015 (2008)
  [arXiv:0810.1563 [hep-th]].

\bibitem{Hartnoll:2009sz}
  S.~A.~Hartnoll,
``Lectures on holographic methods for condensed matter physics,''
 Class.\ Quant.\ Grav.\  {\bf 26}, 224002 (2009)  [arXiv:0903.3246 [hep-th]].

\bibitem{Herzog:2009xv}
  C.~P.~Herzog,
  ``Lectures on Holographic Superfluidity and Superconductivity,''
  J.\ Phys.\ A {\bf 42}, 343001 (2009)
  [arXiv:0904.1975 [hep-th]].

\bibitem{McGreevy:2009xe}
  J.~McGreevy,
  ``Holographic duality with a view toward many-body physics,''
  Adv.\ High Energy Phys.\  {\bf 2010}, 723105 (2010)
  [arXiv:0909.0518 [hep-th]].


\bibitem{Horowitz:2010gk}
  G.~T.~Horowitz,
  ``Introduction to Holographic Superconductors,''
  Lect.\ Notes Phys.\  {\bf 828}, 313 (2011)
  [arXiv:1002.1722 [hep-th]].

\bibitem{Cai:2015cya}
  R.~G.~Cai, L.~Li, L.~F.~Li and R.~Q.~Yang,
  ``Introduction to Holographic Superconductor Models,''
  Sci.\ China Phys.\ Mech.\ Astron.\  {\bf 58}, no. 6, 060401 (2015)
  [arXiv:1502.00437 [hep-th]].

\bibitem{Lee:2008xf}
  S.~S.~Lee,
  ``A Non-Fermi Liquid from a Charged Black Hole: A Critical Fermi Ball,''
  Phys.\ Rev.\ D {\bf 79}, 086006 (2009)
  [arXiv:0809.3402 [hep-th]].

\bibitem{Liu:2009dm}
  H.~Liu, J.~McGreevy and D.~Vegh,
  ``Non-Fermi liquids from holography,''
  Phys.\ Rev.\ D {\bf 83}, 065029 (2011)
  [arXiv:0903.2477 [hep-th]]

\bibitem{Cubrovic:2009ye}
  M.~Cubrovic, J.~Zaanen and K.~Schalm,
  ``String Theory, Quantum Phase Transitions and the Emergent Fermi-Liquid,''
  Science {\bf 325}, 439 (2009)
  [arXiv:0904.1993 [hep-th]].

\bibitem{Aperis:2010cd}
  A.~Aperis, P.~Kotetes, E.~Papantonopoulos, G.~Siopsis, P.~Skamagoulis and G.~Varelogiannis,
  ``Holographic Charge Density Waves,''
  Phys.\ Lett.\ B {\bf 702}, 181 (2011)
  [arXiv:1009.6179 [hep-th]].


\bibitem{Donos:2013gda}
  A.~Donos and J.~P.~Gauntlett,
  ``Holographic charge density waves,''
  Phys.\ Rev.\ D {\bf 87}, 126008 (2013)
  [arXiv:1303.4398 [hep-th]].

\bibitem{Ling:2014saa}
  Y.~Ling, C.~Niu, J.~Wu, Z.~Xian and H.~b.~Zhang,
  ``Metal-insulator Transition by Holographic Charge Density Waves,''
  Phys.\ Rev.\ Lett.\  {\bf 113}, 091602 (2014)
  [arXiv:1404.0777 [hep-th]].


\bibitem{Murata:2010dx}
  K.~Murata, S.~Kinoshita and N.~Tanahashi,
  ``Non-equilibrium Condensation Process in a Holographic Superconductor,''
  JHEP {\bf 1007}, 050 (2010)
  [arXiv:1005.0633 [hep-th]].

\bibitem{Bhaseen:2012gg}
  M.~J.~Bhaseen, J.~P.~Gauntlett, B.~D.~Simons, J.~Sonner and T.~Wiseman,
  ``Holographic Superfluids and the Dynamics of Symmetry Breaking,''
  Phys.\ Rev.\ Lett.\  {\bf 110}, 015301 (2013)
  [arXiv:1207.4194 [hep-th]].

\bibitem{Adams:2012pj}
  A.~Adams, P.~M.~Chesler and H.~Liu,
  ``Holographic Vortex Liquids and Superfluid Turbulence,''
  Science {\bf 341}, 368 (2013)
  [arXiv:1212.0281 [hep-th]].

\bibitem{Garcia-Garcia:2013rha}
  A.~M.~Garc¨ªa-Garc¨ªa, H.~B.~Zeng and H.~Q.~Zhang,
  ``A thermal quench induces spatial inhomogeneities in a holographic superconductor,''
  JHEP {\bf 1407}, 096 (2014)
  [arXiv:1308.5398 [hep-th]].

\bibitem{Chesler:2013lia}
  P.~M.~Chesler and L.~G.~Yaffe,
  ``Numerical solution of gravitational dynamics in asymptotically anti-de Sitter spacetimes,''
  JHEP {\bf 1407}, 086 (2014)
  [arXiv:1309.1439 [hep-th]].

\bibitem{Cai:2014oca}
  R.~G.~Cai and R.~Q.~Yang,
  ``Paramagnetism-Ferromagnetism Phase Transition in a Dyonic Black Hole,''
  Phys.\ Rev.\ D {\bf 90}, no. 8, 081901 (2014)
  [arXiv:1404.2856 [hep-th]].

\bibitem{Cai:2015bsa}
  R.~G.~Cai and R.~Q.~Yang,
  ``Antisymmetric tensor field and spontaneous magnetization in holographic duality,''
  arXiv:1504.00855 [hep-th].

\bibitem{Cai:2015jta}
  R.~G.~Cai, R.~Q.~Yang, Y.~B.~Wu and C.~Y.~Zhang,
  ``Massive $2$-form field and holographic ferromagnetic phase transition,''
  JHEP {\bf 1511}, 021 (2015)
  [arXiv:1507.00546 [hep-th]].



\bibitem{Lonzarich}
 G. G. Lonzarich;S. S. Saxena;P. Agarwal;K. Ahilan;F. M. Grosche;R. K. W. Haselwimmer;M. J. Steiner;E. Pugh, et al. "Superconductivity on the border of itinerant-electron ferromagnetism in UGe2", Nature {\bf 406} 587 (2000).

\bibitem{Aoki}
D. Aoki, A. Huxley, E. Ressouche, D. Braithwaite, J. Flouquet, J-P. Brison, E. Lhotel and C. Paulsen, ``Coexistence of superconductivity and ferromagnetism in URhGe", Nature {\bf 413} 613 (2001):

\bibitem{Huy}
N. T. Huy, A. Gasparini, et al., ``Superconductivity on the border of weak itinerant ferromagnetism in UCoGe", Phys. Rev. Lett.  {\bf 99}, 067006 (2007)

\bibitem{Cai:2013aca}
  R.~G.~Cai, L.~Li and L.~F.~Li,
  ``A Holographic P-wave Superconductor Model,''
  JHEP {\bf 1401}, 032 (2014)
  [arXiv:1309.4877 [hep-th]].

\bibitem{Cai:2014dza}
  R.~G.~Cai and R.~Q.~Yang,
  ``Coexistence and competition of ferromagnetism and $p$-wave superconductivity in holographic model,''
  Phys.\ Rev.\ D {\bf 91}, no. 2, 026001 (2015)
  doi:10.1103/PhysRevD.91.026001
  [arXiv:1410.5080 [hep-th]].

\bibitem{Urushibara}
A. Urushibara, Y. Moritomo, T. Arima, A. Asamitsu, G. Kido, Y. Tokura, ``Insulator-metal transition and giant magnetoresistance in La$_{1-x}$Sr$_x$MnO$_3$", Phys. Rev. B {\bf 51}, 14,103 (1995).

\bibitem{Uehara}
Uehara, M., Mori, S., Chen, C. H., and Cheong, S. W., ``Percolative phase separation underlies colossal magnetoresistance in mixed-valent manganites", Nature, {\bf 399} (6736), 560-563 (1999).

\bibitem{MBS}
M. B. Salamon, and M. Jaime, ``The physics of manganites: Structure and transport",  Rev. Mod. Phys. {\bf 73}, 583 (2001).

\bibitem{Dagottoa}
E. Dagotto, T. Hotta, A. Moreo, ``Colossal magnetoresistant materials: the key role of phase separation", Phy. Rep. {\bf 344}, 1-153 (2001).

\bibitem{Nagaev}
E.L. Nagaev, ``Colossal-magnetoresistance materials: manganites and conventional ferromagnetic semiconductors", Phys. Rep. {\bf 346}, 387-531 (2001).

\bibitem{Cengiz}
Cengiz Sen, Gonzalo Alvarez, Elbio Dagotto, ``Unveiling First Order CMR Transitions in the Two-Orbital Model for Manganites", Phys. Rev. Lett. {\bf 105}, 097203 (2010).

\bibitem{Mukherjee}
Mukherjee Anamitra, Cole William S., Woodward Patrick, Randeria Mohit and Trivedi Nandini, ``Theory of Strain-Controlled Magnetotransport and Stabilization of the Ferromagnetic Insulating Phase in Manganite Thin Films", Phys. Rev. Lett. {\bf 110}, 157201 (2013).

\bibitem{Cai:2015wfa}
  R.~G.~Cai and R.~Q.~Yang,
  ``Insulator/metal phase transition and colossal magnetoresistance in holographic model,''
  Phys.\ Rev.\ D {\bf 92}, no. 10, 106002 (2015)
  [arXiv:1507.03105 [hep-th]].

\bibitem{Cai:2014jta}
  R.~G.~Cai and R.~Q.~Yang,
  ``Holographic model for the paramagnetism/ antiferromagnetism phase transition,''
  Phys.\ Rev.\ D {\bf 91}, no. 8, 086001 (2015)
  [arXiv:1404.7737 [hep-th]].

\bibitem{Cai:2015mja}
  R.~G.~Cai, R.~Q.~Yang and F.~V.~Kusmartsev,
  ``Holographic model for antiferromagnetic quantum phase transition induced by magnetic field,''
  Phys.\ Rev.\ D {\bf 92}, no. 8, 086001 (2015)
  [Phys.\ Rev.\ D {\bf 92}, 086001 (2015)]
  [arXiv:1501.04481 [hep-th]].

\bibitem{Cai:2015xpa}
  R.~G.~Cai, R.~Q.~Yang and F.~V.~Kusmartsev,
  ``Holographic antiferromganetic quantum criticality and AdS$_2$ scaling limit,''
  Phys.\ Rev.\ D {\bf 92}, 046005 (2015)
  [arXiv:1505.03405 [hep-th]].

\bibitem{Niven}
J.F. Niven, et al, ``Magnetic phase transitions and magnetic entropy in the XY antiferromagnetic pyrochlores(Er$_{1-x}$Y$_x$ )$_2$Ti$_2$O$_7$", Proc. R. Soc. A {\bf 470}: 20140387 (2014).
\end{thebibliography}
\end{document}